\documentclass[aps,onecolumn,nofootinbib,superscriptaddress,tightenlines]{revtex4}

\usepackage{amssymb,amstext,amsmath,amsthm,slashed}
\usepackage[dvips]{graphicx}
\usepackage{latexsym}
\usepackage{psfrag}
\usepackage{amsfonts}
\usepackage{bbm} 
\usepackage{color}
\usepackage[all,cmtip]{xy}
\usepackage{verbatim}
\usepackage{bbm}
\usepackage{dcolumn}
\usepackage{tabularx}
\usepackage{leftidx}
\usepackage{tensor}
\usepackage{boxedminipage}
\usepackage{fancyvrb}
\usepackage{float}
\usepackage{wrapfig}
\usepackage{booktabs}
\usepackage{dsfont} 
\usepackage{mathrsfs}
\usepackage{hyperref}

\newcommand{\Proof}{\begin{proof}}
\newcommand{\QED}{\end{proof} \noindent}

\begin{document}

\title{First-order quantum-gravitational correction to Friedmannian cosmology \\ from covariant, holomorphic spinfoam cosmology}

\vskip.9in

\author{Christian R\"oken}

\vskip.9in

\affiliation{Universit\"at Regensburg, Fakult\"at f\"ur Mathematik, 93040 Regensburg, Germany \footnotetext{e-mail: christian.roeken@mathematik.uni-regensburg.de}}
\affiliation{Centre de Physique Th\'{e}orique de Luminy, Case $907$, $13288$ Marseille, France}

\vskip.9in 

\date{\today}

\begin{abstract}
\vspace{0.4cm} \noindent \textbf{\footnotesize ABSTRACT.} \, The first-order loop quantum gravity correction of the simplest, classical general-relativistic Friedmann Hamiltonian constraint, emerging from a holomorphic spinfoam cosmological model peaked on homogeneous, isotropic geometries, is studied. The quantum Hamiltonian constraint, satisfied by the EPRL transition amplitude between the boundary cosmological coherent states, includes a contribution of the order of the Planck constant $\hbar$ that also appears in the corresponding semiclassical symplectic model. The analysis of this term gives a quantum-gravitational correction to the classical Friedmann dynamics of the scale factor yielding a small decelerating expansion (small accelerating contraction) of the universe. The robustness of the physical interpretation is established for arbitrary refinements of the boundary graphs. Also, mathematical equivalences between the semiclassical cosmological model and certain classical fluid and scalar field theories are explored.
\end{abstract}

\setcounter{tocdepth}{2}

\vspace{0.1cm}

\maketitle

\tableofcontents

\section{Introduction}

\noindent Bianchi, Rovelli, and Vidotto have introduced a basic spinfoam model of quantum cosmology [1]. They started their analysis from covariant loop quantum gravity as in [2, 3], and computed the transition amplitude $W$ between holomorphic, coherent quantum states of spacetime peaked on homogeneous, isotropic geometries. The model was truncated from the full quantum gravity Hilbert space $\mathcal{H} = \underset{\Gamma}{\bigoplus} \mathcal{H}_{\Gamma}$, with the sum running over abstract graphs $\Gamma$ (i.e., a set of $L$ links $l$ and $N$ nodes $n$ including two relations $s$ and $t$ assigning source and target nodes to each link), down to a single graph space $\mathcal{H}_{\Gamma} = L_2\bigl[SU(2)^L/SU(2)^N\bigr]$, choosing the graph $\Gamma = \partial \, \mathcal{C}$ to be a dipole graph, and to a first-order vertex expansion of the two-complex $\mathcal{C}$ in the large volume limit, where the scale factor of the universe $R$ is much larger than the Planck length $l_{\textnormal{P}} \sim 10^{-35} \textnormal{m}$. The theory includes all states $\left|\Psi\right\rangle$ that have support on subgraphs of $\Gamma$ since they are already contained in $\mathcal{H}_{\Gamma}$. It loses states that need to be defined on larger graphs. This quantum regime approximation corresponds classically to a truncation of the degrees of freedom in general relativity to a finite level as given by the lowest modes of a mode expansion of the gravitational field on a compact space [4, 5]. Accordingly, the quantum-gravitational field is described by the interactions of a finite number of quanta of spacetime. The transition amplitude $W$ turned out to be in the kernel of a differential operator $\widehat{H}$, i.e., $\widehat{H} W = 0$, of which the classical limit is the Hamiltonian constraint of the flat Friedmannian cosmology in the absence of both a mass-energy content and a cosmological constant, describing the simplest but unphysical universe with a rather trivial dynamics. More precisely, the operator $\widehat{H}$ is the Friedmann operator plus a term proportional to $\hbar$. Here, this term as well as its physical implications and their robustness are studied. It is also shown that its dynamical effect is the same as that of an ultralight, irrotational, stiff, perfect fluid matter content filling a classical universe [6, 7]. Furthermore, a mathematical correspondence between a stiff fluid and a scalar field, as demonstrated in [8], links the quantum correction to an effective, massless, real scalar field.

Of course, the Bianchi-Rovelli-Vidotto (BRV) model depends on the specific assumptions made for the truncation of the degrees of freedom. This work is not meant to clarify, much less to solve the conceptual issues raised by these assumptions. Only the physics of the results of the BRV model is analyzed. Further, it is important to stress that the BRV model includes new degrees of freedom beyond that of standard cosmology because the geometry of the triangulation used in the construction of the two-complex $\mathcal{C}$ captures more degrees of freedom of the classical metric than just the scale factor [4, 5]. While the quantum-gravitational states are peaked on homogeneous and isotropic geometries, inhomogeneous and anisotropic degrees of freedom fluctuating around zero, are also taken into account. Therefore, the effect studied can be viewed either as a result of the quantum theory, or the extra classical degrees of freedom, or both.

\section{Semiclassical Friedmannian Cosmology}

\noindent A semiclassical cosmological theory based on the loop quantum gravity BRV spinfoam model is derived, allowing for an analysis of the emergent physics of a particular spinfoam model of quantum cosmology in an effective setting. The corresponding first-order quantum correction to the classical Friedmann Hamiltonian constraint is found by making use of the fact that the Euclidean EPRL transition amplitude $W$ between different homogeneous, isotropic, coherent quantum-gravitational boundary states is in the kernel of a quantum Hamiltonian constraint operator $\widehat{H}$, which yields a semiclassical phase space formulation in a suitable limit. It is this symplectic model that leads to a first-order quantum-geometry-corrected Friedmann Hamiltonian constraint. The validity and robustness of the results presented in this study are tested afterward within the Lorentzian framework for arbitrary refinements of regular (i.e., all nodes have the same valency), abstract boundary graphs.  

In the EPRL model [2], transition amplitudes $\mathcal{Z}_{\mathcal{C}}$ associated to a given two-complex in the Euclidean theory are defined by
\begin{equation}\label{eprlta}
\mathcal{Z}_{\mathcal{C}} = \sum_{j_f, i_e} \prod_{f} d_f \prod_{v} W_v(j_f, i_e),
\end{equation}
where $d_f = (|1 - \gamma| j_f + 1) ((1 + \gamma) j_f + 1)$, $j_f \in \mathbb{N}/2$ is the spin related to a face $f$, $i_e$ is an intertwiner corresponding to an edge $e$, $W_v$ denotes the Euclidean vertex amplitude of a vertex $v$, and $\gamma \in \mathbb{C}$ is the Barbero-Immirzi parameter. The vertex amplitude assumes a simple form when it is evaluated in the holomorphic Hilbert space basis of the gauge invariant coherent states
\begin{equation*}
\Psi_{H_l}(U_l) = \int_{\textnormal{SU(2)}^N} \prod_{n} \textnormal{d}g_n \prod_{l} K_t\Bigl(g^{-1}_{s(l)} \, U_l \, g_{t(l)} \, H^{-1}_l\Bigr),
\end{equation*}
which are defined as integrals on SU(2) with $U_l \in$ SU(2), $H_l \in$ SL(2,$\mathbb{C}$), and the analytic continuation $K_t$ of the heat kernel function on SU(2) to SL(2,$\mathbb{C}$). Explicitly, the holomorphic representation of the Euclidean vertex amplitude becomes
\begin{equation}\label{ETA}
W_v = \int_{\textnormal{SO(4)}^N} \prod_{n} \textnormal{d}G_n \prod_{l} \sum_{j_l} (2 j_l + 1) \exp{\bigl(- 2 t \hbar j_l (j_l + 1)\bigr)} \, \textnormal{Tr}\Bigl(D^{(j_l)}(H_l) \, Y^{\dagger} \, D^{(j^+_l, \, j^-_l)}\bigl(G_{s(l)} \, G^{-1}_{t(l)}\bigr) \, Y\Bigr),
\end{equation}
where $D^{(j_l)}$ is the analytic continuation of the Wigner matrix of the spin-$j_l$ representation of SU(2) to SL(2,$\mathbb{C}$), $D^{(j^+_l, \, j^-_l)}$ is the Wigner matrix of the spin-$(j^+_l, j^-_l)$ representation of SO(4), and $Y$ is a map gluing these matrices of different dimensions together. In terms of the vertex amplitude in the representation (\ref{ETA}), the transition amplitude (\ref{eprlta}) between two homogeneous, isotropic, coherent states in the BRV cosmological model [1, 9], i.e., formulated on a dipole graph space in the first-order vertex expansion and large volume limit, is given by
\begin{equation} \label{1}
W(z_{\textnormal{i}}, z_{\textnormal{f}}) = N^2 z_{\textnormal{i}} z_{\textnormal{f}} \exp{\Biggl(- \frac{z_{\textnormal{i}}^2 + z_{\textnormal{f}}^2}{2 t \hbar} \Biggr)},
\end{equation}
where the parameter $t$ is a measure for the spread of the Gaussian, $N = - 4 \textnormal{i} \pi^2 N_0/(t^3 \hbar)$ is a constant factor which is derived in detail in [1, 12], and the $z_j$ are two complex variables with $j \in \{\textnormal{i} = \textnormal{initial}, \textnormal{f} = \textnormal{final}\}$ reading
\begin{equation*}
z_j = \alpha c_j + \textnormal{i} \beta |p_j|.
\end{equation*}
The functions $c_j = \gamma \dot{R}_j$ and $|p_j| = R_j^2$ are the canonical loop quantum cosmology variables that depend on the standard cosmology scale factor $R_j$ and its derivative $\dot{R}_j$ with respect to a cosmological time parameter [10, 11], and the constants $\alpha$ and $\beta$, or rather their product, is going to be fixed in the course of this section. The transition amplitude (\ref{1}) can be factorized  
\begin{equation*}
W(z_{\textnormal{i}}, z_{\textnormal{f}}) = W_{\textnormal{i}}(z_{\textnormal{i}}) \, W_{\textnormal{f}}(z_{\textnormal{f}})
\end{equation*}
with the functions
\begin{equation*}
W_j(z_j) = N z_j \exp{\Biggl(- \frac{z_j^2}{2 t \hbar}\Biggr)}.
\end{equation*}
These fulfill the (stationary) Schr\"odinger-like operator equation
\begin{equation}\label{4}
\widehat{H} W_j := \Biggl(z_j^2 - t^2 \hbar^2 \, \frac{\partial^2 }{\partial z_j^2} - 3 t \hbar \Biggr) W_j = 0,
\end{equation}
which constitutes an evolution equation of the function $W_j$. The index $j$ is, from now on, suppressed for convenience. The semiclassical limit of this quantum constraint can be found using the following heuristic method. 

A classical dynamical system can be described by a triple $(\mathcal{M}, \boldsymbol{\omega}, H)$, where $\mathcal{M}$ is a $2 m$-dimensional (phase) space with coordinates $\mathfrak{q}^i$ and momenta $\mathfrak{p}_i$, $i = 1, ..., m$, $\boldsymbol{\omega} = \textnormal{d}\mathfrak{p}_i \wedge \textnormal{d}\mathfrak{q}^i$ denotes a symplectic two-form, and $H$ a Hamiltonian. Considering a one-dimensional system with configuration variable $\mathfrak{q} = z$ and conjugated momentum $\mathfrak{p} = \overline{z}/(\textnormal{i} t)$, one obtains the symplectic structure  
\begin{equation}\label{symplstr}
\boldsymbol{\omega} = \frac{\textnormal{i}}{t} \, \textnormal{d}z \wedge \textnormal{d}\overline{z} \,\,\,\,\, \textnormal{and} \,\,\,\,\, \{z, \overline{z}\} = \textnormal{i} t.
\end{equation}
The corresponding quantum theory, satisfying the fundamental commutator 
\begin{equation*}
\bigl[\widehat{z}, \widehat{\overline{z}} \, \bigr] = \textnormal{i} \hbar \{z, \overline{z}\},
\end{equation*}
is, therefore, given by the correspondence principle
\begin{equation*}
z \rightarrow \widehat{z} = z \,\,\,\,\, \textnormal{and} \,\,\,\,\, \overline{z} \rightarrow \widehat{\overline{z}} = t \hbar \frac{\partial}{\partial z}.
\end{equation*}
In this representation, $\widehat{z}$ and $\widehat{\overline{z}}$ are multiplicative and derivative operators, respectively. By means of this correspondence principle, a transition from the quantum constraint equation (\ref{4}) to a semiclassical, symplectic theory of the form
\begin{equation*}
z^2 - \overline{z}^2 - 3 t \hbar = 0
\end{equation*}
can be realized. In terms of the scale factor and its derivative, this equation leads to a first-order quantum-geometry-corrected, semiclassical Friedmann Hamiltonian constraint
\begin{equation}\label{5}
H_{\textnormal{Fr.}} = H_{\textnormal{Cl.}} + H_{\textnormal{Qu.}} = - \frac{3}{8 \pi G} \, \Biggl(R \dot{R}^2 + \frac{3 \textnormal{i} t \hbar}{4 \alpha \beta \gamma} \, \frac{\dot{R}}{R} \Biggr)= 0,
\end{equation}
where $H_{\textnormal{Cl.}} = - 3 R \dot{R}^2/(8 \pi G)$ is the Hamiltonian constraint that governs the gravitational part of the dynamics of classical general-relativistic Friedmannian cosmology. Note that since the precise relation between the transition amplitudes of the EPRL model and any canonical loop quantum gravity dynamics is unknown, nonetheless there are certain points of contact, the quantum Hamiltonian constraint (\ref{4}) and the symplectic structure (\ref{symplstr}) have to be taken as inputs. Furthermore, after a derivation of the EPRL transition amplitude in a coordinate-independent setting, a change from generally covariant variables to coordinates in a Hamiltonian formulation was achieved by formally parameterizing the scale factor and its derivative in terms of a cosmological time $T$. With all due caveats, the specific first-order BRV quantum contribution $H_{\textnormal{Qu.}}$ to the classical Friedmann constraint $H_{\textnormal{Cl.}}$ is, in the following, taken seriously, and the physics it describes is analyzed. 

The product $\alpha \beta$ appearing in $H_{\textnormal{Qu.}}$ can be fixed by relating the Poisson bracket of $z$ and $\overline{z}$ in (\ref{symplstr}) to the Poisson bracket of the loop quantum cosmology variables $c$ and $|p|$ [10], namely 
\begin{equation*}
\{z, \overline{z}\} = - 2 \textnormal{i} \alpha \beta \{c, |p|\} = - \frac{16 \textnormal{i} \alpha \beta \pi \gamma G}{3}, 
\end{equation*}
yielding 
\begin{equation*}
\alpha \beta = - \frac{3 t}{16 \pi \gamma G} \,\, \in \mathbb{R} 
\end{equation*}
for real values of the Barbero-Immirzi parameter. Then, for $\textnormal{d}R/\textnormal{d}T \not= 0$, the Hamiltonian constraint (\ref{5}) can be recast as 
\begin{equation}\label{esfe}
\frac{\textnormal{d}R}{\textnormal{d}T} - \frac{4 \pi \textnormal{i} l_{\textnormal{P}}^2}{R^2} = 0 \, ,
\end{equation}
where $l_{\textnormal{P}} = \sqrt{\hbar G}$ is the Planck length. (Note that for $\textnormal{d}R/\textnormal{d}T = 0$, one immediately finds both $H_{\textnormal{Cl.}} = 0$ and $H_{\textnormal{Qu.}} = 0$ (cf. Eq.(\ref{5})), leading to the solution $R = \textnormal{const.}$ from which one concludes that in this case no quantum correction to the classical scale factor exists.) Since there is no unique way to select a proper cosmological time for the parameterization of the scale factor, the canonical parameter choices for the cosmological time are discussed.

First, consider a real-valued cosmological time $T \in \mathbb{R}_{\geq 0}$. The solution of the first-order quantum Friedmann Hamiltonian constraint (\ref{esfe}) is easily found by separation of variables and integration with respect to the cosmological time $T$, yielding the solution
\begin{equation}\label{rsf}
R(T) = \bigl(12 \pi \textnormal{i} l_{\textnormal{P}}^2 \, T + \lambda_0\bigr)^{1/3} \in \mathbb{C},
\end{equation}
where $\lambda_0$ denotes a constant. This complex-valued solution is unphysical because a meaningful scale factor is a positive real-valued function and, thus, this solution has to be disregarded. For the remaining, imaginary choices of the cosmological time $T = \pm \textnormal{i} \mathcal{T} \in \mathbb{C}$ with $\mathcal{T} \in \mathbb{R}_{\geq 0}$, the corresponding first-order quantum Friedmann Hamiltonian constraints are given by the differential equations 
\begin{equation}
\frac{\textnormal{d}R}{\textnormal{d}\mathcal{T}} \pm \frac{4 \pi l_{\textnormal{P}}^2}{R^2} = 0,
\end{equation}
which can be solved by the same method as in the latter case. These solutions read
\begin{equation}\label{csf}
R(\mathcal{T}) = \bigl(\mp 12 \pi l_{\textnormal{P}}^2 \, \mathcal{T} + \lambda_0\bigr)^{1/3},
\end{equation}
describing either growing universes with small decelerating expansions of the order $l_{\textnormal{P}}^{2/3}$ (positive solution) or shrinking universes with accelerating contractions of the same order (negative solution) with initial values at $R = \lambda_0^{1/3}$ and cutoffs for the shrinking universe solutions at $R = 0$. It is sufficient to study only the expanding universe scenario because the shrinking universe scenario can be obtained by a discrete time reflexion transformation $\mathcal{T} \mapsto - \mathcal{T}$ applied to the latter. Hence, in the subsequent sections, only the positive solution is analyzed. The undetermined constant $\lambda_0$ can be fixed by imposing initial conditions on the cosmology. Further, it should be noted that the semiclassical scale factor does not depend on the parameter $t$ which is originally used in the coherent states. It is not entering the final result because the dependence on $t$ is ``absorbed'' into the symplectic structure (\ref{symplstr}) due to the specific choice of the conjugated momentum $\mathfrak{p} = \overline{z}/(\textnormal{i} t)$. In the classical limit $l_{\textnormal{P}} \searrow 0$, one recovers the well-known result $R(T) = \lambda_0^{1/3} = \textnormal{const.}$ from cosmological studies in the context of pure gravity without matter which is, in the large volume limit, just flat space. Since the BRV quantum-cosmological model includes fluctuations of the inhomogeneous and anisotropic degrees of freedom, the scale factor modification presented in (\ref{csf}) can either be seen as being of quantum origin, or to be due to the extra degrees of freedom introduced in the BRV model, or both.

\section{Robustness of the Semiclassical Cosmological Model}

\noindent By appropriately including higher-order contributions in the graph and vertex expansions to improve the accuracy of the truncations, one would naturally expect to obtain changes regarding the transition amplitude and, thus, for the semiclassical model. However, [13] provides a procedure within the Lorentzian EPRL spinfoam model to recover more degrees of freedom by refining the complex, indicating that the form of the quantum-gravitational correction in the semiclassical limit does not depend on those refinements of the boundary graphs. 

More precisely, the BRV cosmological model was generalized in the Lorentzian EPRL model with spinfoam transition amplitudes 
\begin{equation*}
\mathcal{Z}_{\mathcal{C}} = \sum_{j_f, i_e} \prod_{f} (2 j_f + 1) \prod_{v} W_v(j_f, i_e)
\end{equation*}
for any regular, abstract boundary graph with arbitrary numbers of nodes $N$ and links $L$. The Lorentzian vertex amplitude $W_v$ is defined as an integral over SL(2,$\mathbb{C}$). The first-order vertex expansion of the transition amplitude between homogeneous, isotropic, coherent states in the large volume limit reads in this setting   
\begin{equation}\label{GLTA}
W_{L, N}(z) = N_{\Gamma}(L, N) \, \biggl(\frac{4 \pi}{t}\biggr)^{L/2} \, \biggl(\frac{z}{4 \textnormal{i} t \hbar}\biggr)^{L - 3}  \, \exp{\biggl(- \frac{L}{8 t \hbar} \, z^2\biggr)},
\end{equation}
which is in general of a different functional form than the much simpler Euclidean EPRL transition amplitude (\ref{1}), as expected when introducing new degrees of freedom, namely $L$ and $N$. Note that applying the same rudimentary dipole graph expansion with $L = 4$ and $N = 2$ in the Lorentzian framework leads to exactly the same result as presented within the Euclidean model. Hence, no differences of the functional form occur due to the change from the Euclidean to the Lorentzian model. The transition amplitude (\ref{GLTA}) satisfies the Schr\"odinger-like quantum constraint equation
\begin{equation*}
\widehat{H} \, W_{L, N}(z) := \biggl(z^2 - \frac{16 t^2 \hbar^2}{L^2} \, \frac{\partial^2}{\partial z^2} - \frac{4 t \hbar (2 L - 5)}{L} + \frac{16 t^2 \hbar^2 (L - 4) (L - 3)}{L^2 z^2}\biggr) W_{L, N}(z) = 0.
\end{equation*}
Restricting the analysis to first-order quantum-gravitational corrections in the large scale limit, and using the correspondence principle 
\begin{equation*}
z \rightarrow \widehat{z} = z \,\,\,\,\, \textnormal{and} \,\,\,\,\, \overline{z} \rightarrow \widehat{\overline{z}} = \frac{4 t \hbar}{L} \frac{\partial}{\partial z},
\end{equation*}
one obtains the following semiclassical equation
\begin{equation*}
z^2 - \overline{z}^2 - \frac{4 t \hbar (2 L - 5)}{L} = 0,
\end{equation*}
which, in terms of the scale factor (with parametrization $T = - \textnormal{i} \mathcal{T}$) and its derivative, is given by
\begin{equation}\label{GSCCE}
\frac{\textnormal{d}R}{\textnormal{d}\mathcal{T}} - \frac{16 \pi l_{\textnormal{P}}^2 (2 L - 5)}{3 L R^2} = 0.
\end{equation}
The solution to this differential equation yields
\begin{equation}\label{GSF}
R(\mathcal{T}) = \biggl(\frac{16 \pi l_{\textnormal{P}}^2 (2 L - 5)}{L} \, \mathcal{T} + \lambda_0 \biggr)^{1/3}.
\end{equation}
The extended effective dynamics as stated above for arbitrary refinements of the boundary graphs shows the control of the behavior the semiclassical theory under such changes of the two-complex, indicating the robustness of the physical interpretation at least with respect to the specific vertex expansion chosen. The semiclassical behavior of the generalized Lorentzian model is, thus, identical to the one emerging from the Euclidean BRV model and is independent of the (regular) boundary graph structures.

\section{Semiclassical Cosmological Model, Ultralight, Irrotational, Stiff, Perfect Fluid, and Massless, Real Scalar Field Theory}

\noindent An isotropic expansion of space as described by the scale factor (\ref{GSF}) can also be obtained in a classical setting considering a Friedmannian cosmology in the presence of a specific fluid matter content. Thus, a comparison of the temporal evolutions of the semiclassical scale factor with a classical scale factor for a universe with this mass-energy content is presented. In particular, it is shown that the quantum-geometrical correction resembles the effect of an ultralight, irrotational, stiff, perfect fluid in a classical, flat Friedmann universe in the absence of a cosmological constant. 

In order to study the cosmological dynamics of such a perfect fluid in a homogeneous and isotropic spacetime, one solves the first Friedmann equation and the equation for mass-energy conservation (cf. Eqs.(\ref{FRE}) and (\ref{cons})) simultaneously. These dynamical equations are usually derived in all generality in a flat Minkowskian background using the metric $(\eta^{\mu \nu}) = \textnormal{diag}(-1, 1, 1, 1)$ as well as the energy-momentum tensor for a perfect fluid 
\begin{equation}\label{emt}
T^{\mu \nu} = \bigl(\rho + P\bigr) u^{\mu} u^{\nu} + P \, \eta^{\mu \nu}.
\end{equation}
The quantity $\rho$ is the energy density, $P$ is the isotropic pressure and $(u^{\mu}) = (1, 0, 0, 0)$ is the fluid's four-velocity in a comoving inertial frame. Then, the $(0, 0)$-component and the trace of the Einstein equations are leading to the independent Friedmann equations 
\begin{equation}\label{FRE}
\biggl(\frac{1}{R} \, \frac{\textnormal{d}R}{\textnormal{d}\mathcal{T}}\biggr)^2 = \frac{8 \pi G}{3} \, \rho 
\end{equation}
and
\begin{equation}\label{SCFRE}
\frac{1}{R} \, \frac{\textnormal{d}^2R}{\textnormal{d}\mathcal{T}^2} = - \frac{4 \pi G}{3} (\rho + 3 P),
\end{equation}
respectively. The equation for mass-energy conservation can be obtained by differentiating Eq.(\ref{FRE}) with respect to the time parameter and substituting the result into the second Friedmann equation (\ref{SCFRE}), yielding
\begin{equation}\label{cons}
\frac{\textnormal{d}}{\textnormal{d}\mathcal{T}}\bigl(\rho R^3\bigr) = - P \frac{\textnormal{d}R^3}{\textnormal{d}\mathcal{T}}.
\end{equation}
In the special case of a stable, ultralight, irrotational, stiff, perfect fluid being present as matter content [6, 7], the general perfect fluid equation of state 
\begin{equation*}
P = w \rho
\end{equation*}
becomes 
\begin{equation}\label{eos}
P = \rho
\end{equation}
with an extremal equation of state parameter $w = 1$. This is the largest value consistent with causality, e.g., the speed of sound of the fluid equals the speed of light. The energy density as a function of the scale factor, as implied by the mass-energy conservation equation (\ref{cons}) for such an exotic fluid, is given by
\begin{equation*}
\rho = \frac{\rho_0}{R^6},
\end{equation*}
where $\rho_0$ is a constant. Reinserting this into Eq.(\ref{FRE}) leads after some basic algebraic manipulations to the following differential equation  
\begin{equation*}
\frac{\textnormal{d}R}{\textnormal{d}\mathcal{T}} - \sqrt{\frac{8 \pi G \rho_0}{3}} \,\, \frac{1}{R^2} = 0.
\end{equation*}
This equation matches the form of the semiclassical Hamiltonian constraint equation (\ref{GSCCE}) and, hence, reveals a similarity between the quantum-gravitational contribution $H_{\textnormal{Qu.}}$ and the presence of an ultralight, irrotational, stiff, perfect fluid with $\rho_0 = 32 \pi l_{\textnormal{P}}^4 (2 L - 5)^2/(3 L^2 G)$ in a flat Friedmannian cosmology without cosmological constant. It is noteworthy that this particular fluid corresponds to a massless, real scalar field (for a detailed discussion of the local correspondence between this fluid and the scalar field see [8] and references therein) which can be shown as follows. 

One begins with combining the energy-momentum tensor (\ref{emt}) and the extremal equation of state (\ref{eos}), yielding
\begin{equation}\label{emt2}
T^{\mu \nu} = \rho \bigl(2 u^{\mu} u^{\nu} + \eta^{\mu \nu}\bigr). 
\end{equation}
Since the four-velocity $\boldsymbol{u}$ is irrotational, i.e.,  
\begin{equation*}
\nabla_{[\mu} u_{\nu]} = 0,
\end{equation*}
it can be written in terms of a future-pointing, timelike gradient of a real scalar field
\begin{equation*}
\sqrt{2 \rho} \, u_{\mu} = \nabla_{\mu} \psi.
\end{equation*}
Then, the energy-momentum tensor (\ref{emt2}) becomes
\begin{equation*}
T^{\mu \nu} = \nabla^{\mu} \psi \nabla^{\nu} \psi - \frac{1}{2} \, \eta^{\mu \nu} \nabla_{\sigma} \psi \nabla^{\sigma} \psi.
\end{equation*}
Meeting the general-relativistic energy-momentum conservation condition $\nabla_{\mu} T^{\mu \nu} = 0$, the scalar field $\psi$ satisfies a massless, minimally coupled wave equation
\begin{equation*}
\nabla_{\mu} \nabla^{\mu} \psi = 0.
\end{equation*}
While the equivalences presented here are mathematical similarities between the semiclassical cosmological model, the fluid picture, and the scalar field theory, merely demonstrating that different matter couplings and quantum-geometrical contributions can have the same physical effects, the resemblance between the semiclassical Friedmannian cosmology and the scalar field model in particular strengthens nonetheless the position that a scalar field can be interpreted, in the sense of an effective theory, as a semiclassical approximation to some more general quantum field given, in this context, by the cosmological BRV model.

\section{Summary and Outlook}

\noindent Spinfoam cosmology is a subject of increasing interest. The work done so far has successfully highlighted a cosmological sector in the covariant loop quantization of general relativity by computing transition amplitudes between quantum-gravitational states that describe the geometrical structure of the universe induced by pure gravity without matter. Disregarding the conceptual issues that spinfoam models and, therefore, spinfoam cosmology are facing today, the results obtained from the BRV quantum-cosmological model were considered to hold in order to study the emergent physics in a semiclassical framework. The leading-order quantum-gravitational correction of the order of the Planck constant $\hbar$ was added to the simplest classical Friedmann Hamiltonian constraint. The solution of this particular semiclassical Friedmann model describes a universe with a small decelerating expansion of the order $l_{\textnormal{P}}^{2/3}$ (or a universe with an accelerating contraction of the same order). It was shown that these findings correspond, on the one hand, to those of a classical, flat Friedmannian cosmology without cosmological constant, coupled to an ultralight, irrotational, stiff, perfect fluid matter distribution, and on the other hand, to a massless, real scalar field theory. The quantum-gravitational contribution to the classical Friedmann Hamiltonian constraint and its effect yield the same findings in the frameworks of the Euclidean and the Lorentzian EPRL models. It is robust with respect to changes in the numbers of nodes $N$ and links $L$ of regular, abstract boundary graphs. However, improved spinfoam amplitudes have to be computed at least to the next order in the vertex expansion in order to have irrefutable and conclusive proof of the validity of the approximation and its physical interpretation. 

Further studies of spinfoam models with quantum matter sources are necessary to describe realistic scenarios in quantum cosmology. Applying a semiclassical approximation scheme as presented here may lead to the determination of more precise theoretical limits for astrophysical quantities, like the Hubble parameter, for comparisons with experiments and to an explanation of the concept of dark energy by means of loop quantum gravity modifications. Also, it would be interesting to investigate potential connections of the semiclassical cosmological model, especially the fluid equivalence, to superfluid vacuum theories, an approach in theoretical physics, where the vacuum is considered to be a superfluid.

\section*{Acknowledgments}

\noindent The author is grateful to Carlo Rovelli and Eugenio Bianchi for useful discussions and comments, and to Katharina Proksch and Horst Fichtner for a careful reading of this paper. This work was supported by a DFG Research Fellowship.

\end{document}